\documentclass{tmex20}
\def\Journal#1#2#3#4{{#1} {\bf #2}, #3 (#4)}

\def\AA{\em A\&A}
\def\APJ{\em ApJ}
\def\APJL{\em ApJL}
\def\APH{\em APh}

\def\PRL{\em Phys. Rev. Lett.}

\def\SCIENCE{\em Science}

\def\be{\begin{equation}}
\def\ee{\end{equation}}
\def\bea{\begin{eqnarray}}
\def\eea{\end{eqnarray}}

\newcommand{\Photo}{\includegraphics[height=35mm]{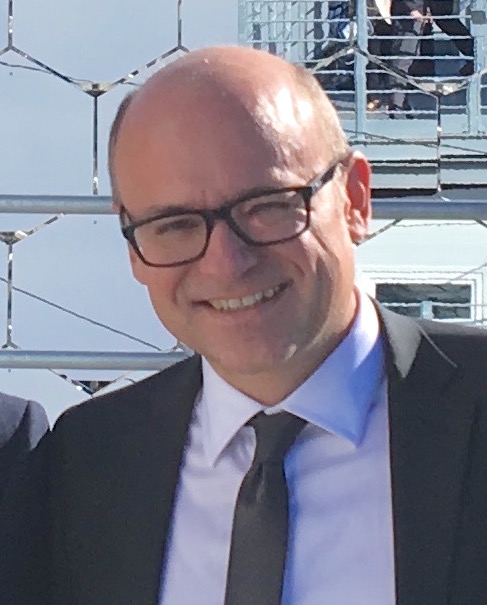}}

\begin{document}
\vspace*{4cm}
\title{THE CHERENKOV TELESCOPE ARRAY}

\author{J. KN\"ODLSEDER (FOR THE CTA CONSORTIUM)}

\address{Institut de Recherche en Astrophysique et Plan\'etologie, 9, avenue Colonel-Roche, \\
31028 Toulouse Cedex 4, France}

\maketitle\abstracts{
The Cherenkov Telescope Array (CTA) is the observatory for ground-based gamma-ray astronomy
that will shape the domain of TeV astronomy for the next decades.
CTA will comprise more than 100 imaging air Cherenkov telescopes deployed on two sites,
one in the northern hemisphere on La Palma and one in the southern hemisphere in Chile.
A large fraction of CTA's observing time will be apportioned through a competitive proposal-driven
time allocation scheme that is open to the scientific community.
Hence CTA will become an astronomical tool that complements other large ground- and space-based
observatories that do and will exist at other wavelengths and for other messengers.
In this contribution I will present the CTA Observatory, its main characteristics, and the current status
of the construction project.
First light was already obtained by prototype CTA telescopes and cameras, and deployment
of the CTA arrays will start soon.
The science analysis software is already in good shape, and available to the wider community for
preparing CTA science and data analysis.}

\section{Introduction}

Our understanding of the Universe has always progressed thanks to the invention of new astronomical
instruments and tools.
This was true for the ancient Greek astronomers, when Galileo turned his telescope towards the sky,
and when new windows of the electromagnetic spectrum were progressively opened after World War II.
Gamma rays were first seen from the sky in 1958, when Peterson and Winckler discovered a burst
of 200-500 keV photons from the Sun during a balloon flight in Cuba.\cite{peterson1958}
Gamma rays with energies above 50 MeV were first detected by the {\em Explorer XI}
satellite in 1961,\cite{kraushaar1962} and TeV gamma rays, which were the most elusive, were
convincingly detected in 1989, when Trevor Weekes and collaborators announced the detection of
$>0.7$~TeV gamma rays from the Crab Nebula using the Imaging Air Cherenkov Telescope (IACT)
Whipple.\cite{weekes1989}

Since then, TeV astronomy has expanded tremendously, thanks to the advent of a new generation
of IACTs comprised of H.E.S.S., MAGIC and VERITAS.
As of today, there are 227 sources of TeV gamma rays listed in the TeVCat catalogue,\footnote{\url{http://tevcat.uchicago.edu}}
and the inventory continues to grow.\cite{wakely2008}
The big surprise of TeV astronomy was the ubiquity of gamma-ray sources.
TeV gamma rays were discovered from Galactic sources such as supernova remnants (SNRs), pulsars
and their wind nebulae, binaries hosting a compact object, and the interstellar medium.
They are also seen in a large variety of extragalactic sources, such as starburst galaxies, radio galaxies,
active galactic nuclei (AGN) and, more recently, also gamma-ray bursts.
All these sources have in common that they accelerate particles to relativistic energies, dominantly
protons and electrons, that can be distinguished through the gamma-ray imprints left by their
interactions with the interstellar medium.

Instrumentation for TeV astronomy is continuing to expand.
Recently, Water Cherenkov Detector Arrays (WCDAs) have complemented the IACTs, with HAWC
inaugurated in 2015 in Mexico and LHAASO being currently deployed in China.
In contrast to IACTs, which have narrow fields-of-view of a few degrees and generally operate only during
moonless nights, WCDAs can operate 24 hours a day and have fields-of-view of about one steradian,
covering a substantial fraction of the sky during one day.
They are hence ideal survey instruments, and excellent tools to unveil extended gamma-ray
sources.\cite{abdo2009,abeysekara2017}
Yet their sensitivity and angular resolution are limited, and to progress in our understanding of the
TeV Universe and the nature of cosmic particle accelerators, a new observatory for ground-based
gamma-ray astronomy is needed.

This new observatory is the Cherenkov Telescope Array (CTA) that is currently developed by a
world-wide partnership of more than 200 institutes.
CTA will be the largest ground-based gamma-ray detection observatory in the world, with more than
100 telescopes in the northern and southern hemispheres.
I will present in this contribution the CTA Observatory, describe its main characteristics, and report
about the development status.
We are not very far from seeing CTA deployed, and only a few years separate us from the exciting
discoveries that come along with a new astronomical instrument.

\section{The CTA Observatory}

CTA will detect gamma rays through the well-established technique of imaging the Cherenkov
light emitted by relativistic electromagnetic particle showers that are created by the interaction of a
gamma ray with the Earth's atmosphere.
These showers produce a Cherenkov light pool with a typical radius of 100-150 metres that 
existing IACTs with their few telescopes sample only sparsely.
With its many telescopes, CTA will determine the characteristics of each particle shower much
more precisely than current IACTs, enhancing the capability to reduce the hadronic particle background,
increasing the angular and energy resolutions, and reducing the trigger threshold of the array.
More telescopes also imply a larger effective detection area.

Consequently, CTA will expand in performance with respect to existing IACTs in all parameters.
CTA will provide a very broad energy coverage, reaching from about 20 GeV up to 300 TeV.
The detection sensitivity will reach about $1/1000$ of the Crab Nebula flux within 50 hours of
observing time, providing access to populations of TeV sources across the entire Galaxy,
and allowing for sampling of AGN variability on a sub-minute time scale. 
The field-of-view will exceed $8^\circ$, enabling the detection of extended and diffuse sources
of gamma-ray emission, and allowing for an effective survey of large fields of the sky.
An angular resolution of about $1$ arcmin will be achieved at the highest energies, providing
the resolution that is needed to resolve SNRs and nearby starburst galaxies.

\begin{figure}
\centerline{\includegraphics[width=\linewidth]{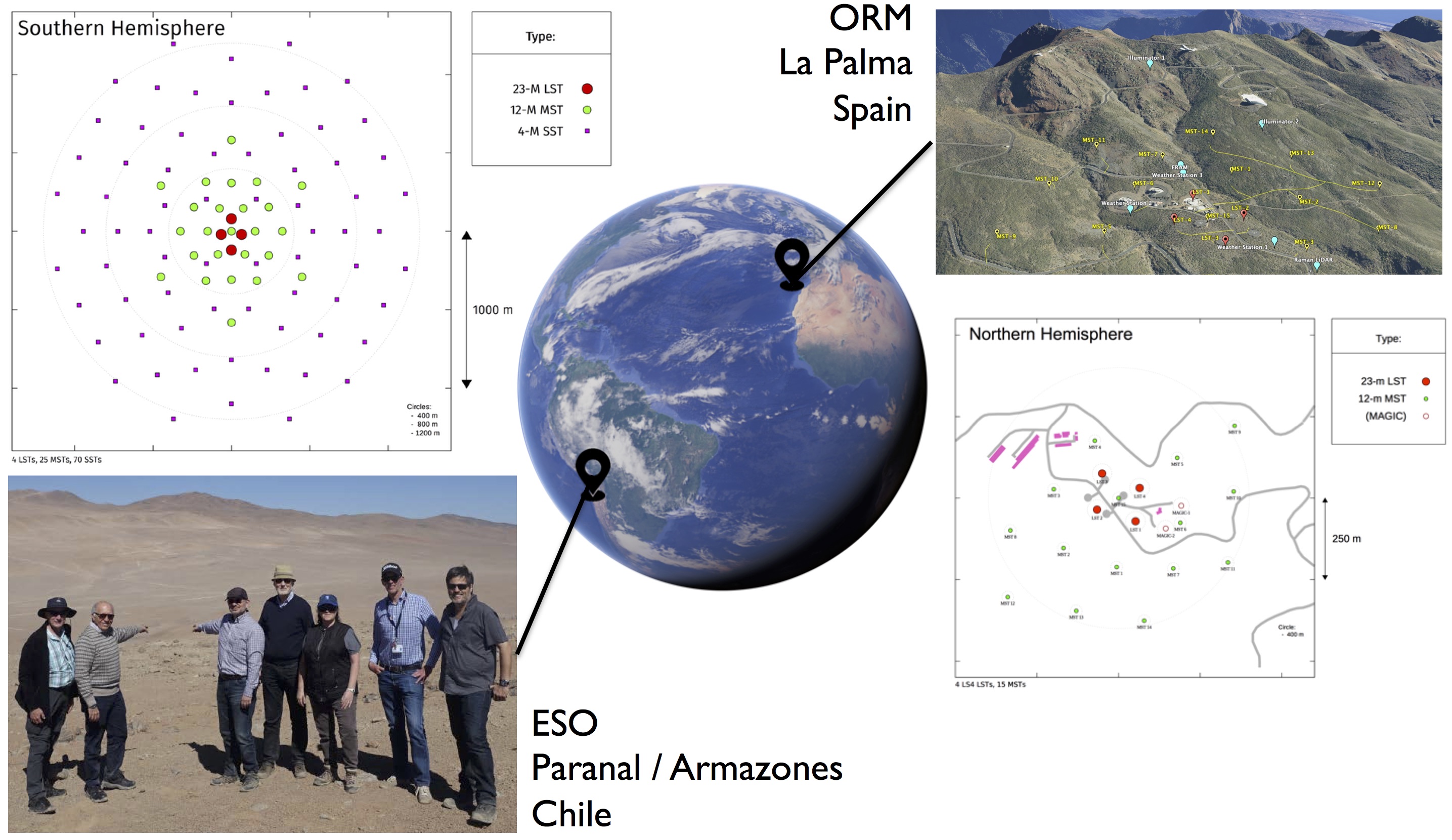}}
\caption[]{
The two telescope array sites of CTA.
The southern site is located near ESO's Paranal Observatory in the Atacama Desert in Chile (left).
The northern site is located at Observatorio del Roque de los Muchachos in Villa de Garafia on the
island of La Palma, Spain (right).}
\label{fig:sites}
\end{figure}

CTA will be comprised of two telescope array sites, one in the southern hemisphere in Chile and
the other in the northern hemisphere on the island of La Palma, Spain (see Fig.~\ref{fig:sites}).
This gives CTA access to the entire sky, allowing for studying rare objects that may not be accessible
from a single site.
While both array sites will be able to observe parts of the Milky Way, the rich inner Galaxy and
in particular the Galactic centre will only be accessible from the southern site.
The extragalactic sky will be visible from both sites, but since it is anticipated that the southern
site will be used more frequently to explore our Galaxy, the northern site will likely more often
scrutinise the extragalactic sky.

\begin{figure}
\centerline{\includegraphics[width=\linewidth]{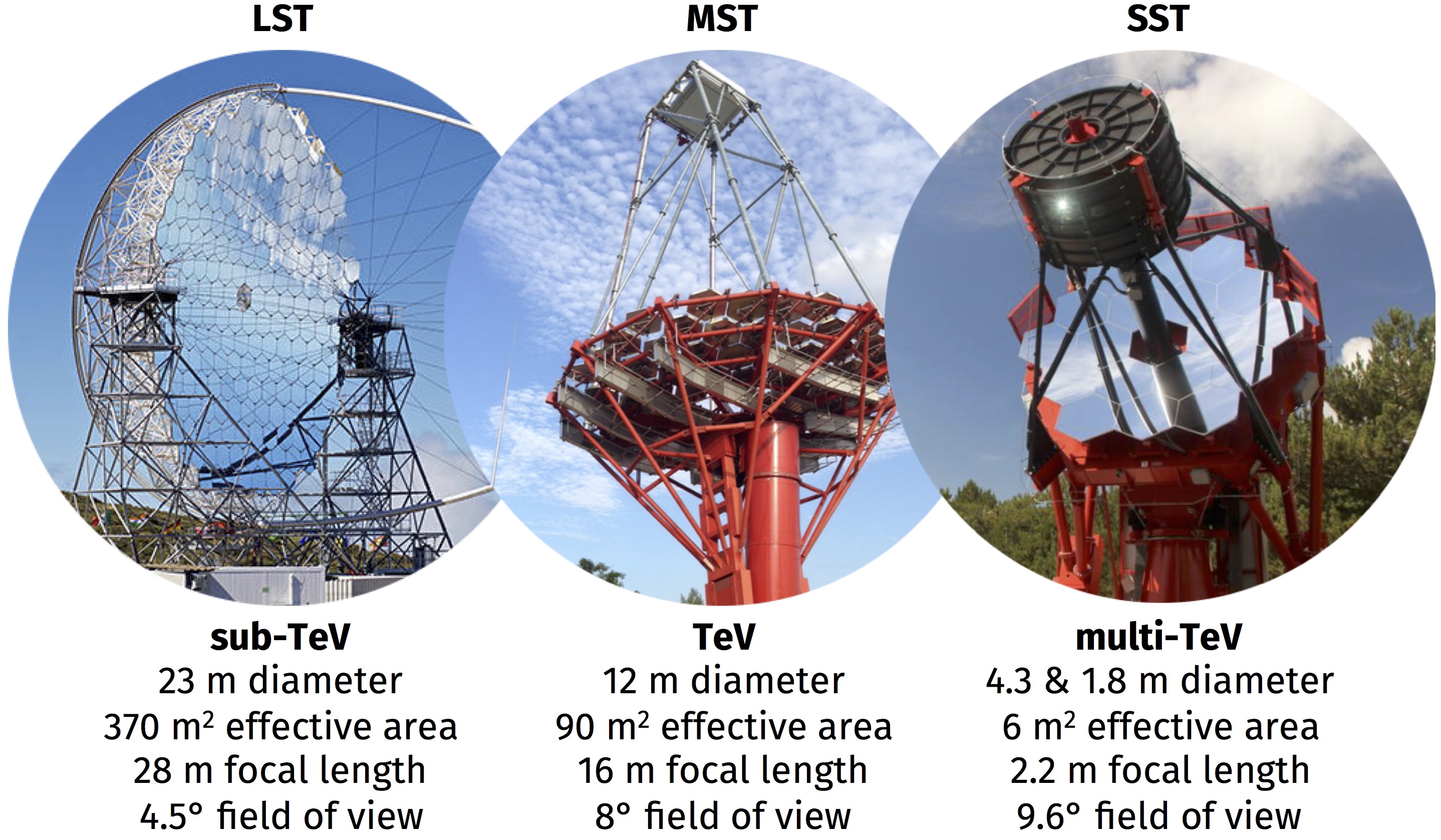}}
\caption[]{
The three telescope types of CTA. The Large-Sized Telescope (LST; left) is optimised for sub-TeV
energies, the Mid-Sized Telescope (MST; centre) is optimised for the TeV range, and the Small-Sized
Telescope (SST; right) is optimised for multi-TeV gamma-ray photons.}
\label{fig:telescopes}
\end{figure}

To cover its broad energy range cost-efficiently, CTA will comprise three types of telescopes.
At the lowest energies, gamma-ray photons are abundant but their Cherenkov light flashes are faint,
hence, a small number of telescopes with large light-collection areas is sufficient.
Conversely, at the highest energies, gamma-ray photons are rare but their Cherenkov light flashes
are bright, hence many telescopes are needed to cover a large detection area, but their light collection
area can be small.
This leads to a design where the sub-TeV range will be observed by a few Large-Sized Telescopes
(LSTs), the TeV range by tens of Mid-Sized Telescopes (MSTs), and the multi-TeV range by a large number
of Small-Sized Telescopes (SSTs; see Fig.~\ref{fig:telescopes}).
The exact CTA baseline layout of the different telescope types can be seen in Fig.~\ref{fig:sites}, with
4 LSTs, 40 MSTs and 70 SSTs in the south, and
4 LSTs and 15 MSTs in the north (no SSTs will be deployed in the north, since multi-TeV gamma
rays are strongly absorbed for extragalactic sources by the cosmic background light).\cite{acharyya2019}
Of course, the energy ranges of the three telescope types will considerably overlap, and many
showers will be seen simultaneously by different telescope types.
The full CTA performance is thus the result of the combination of the three telescope types into
a unique observatory.

\begin{figure}
\centerline{\includegraphics[width=\linewidth]{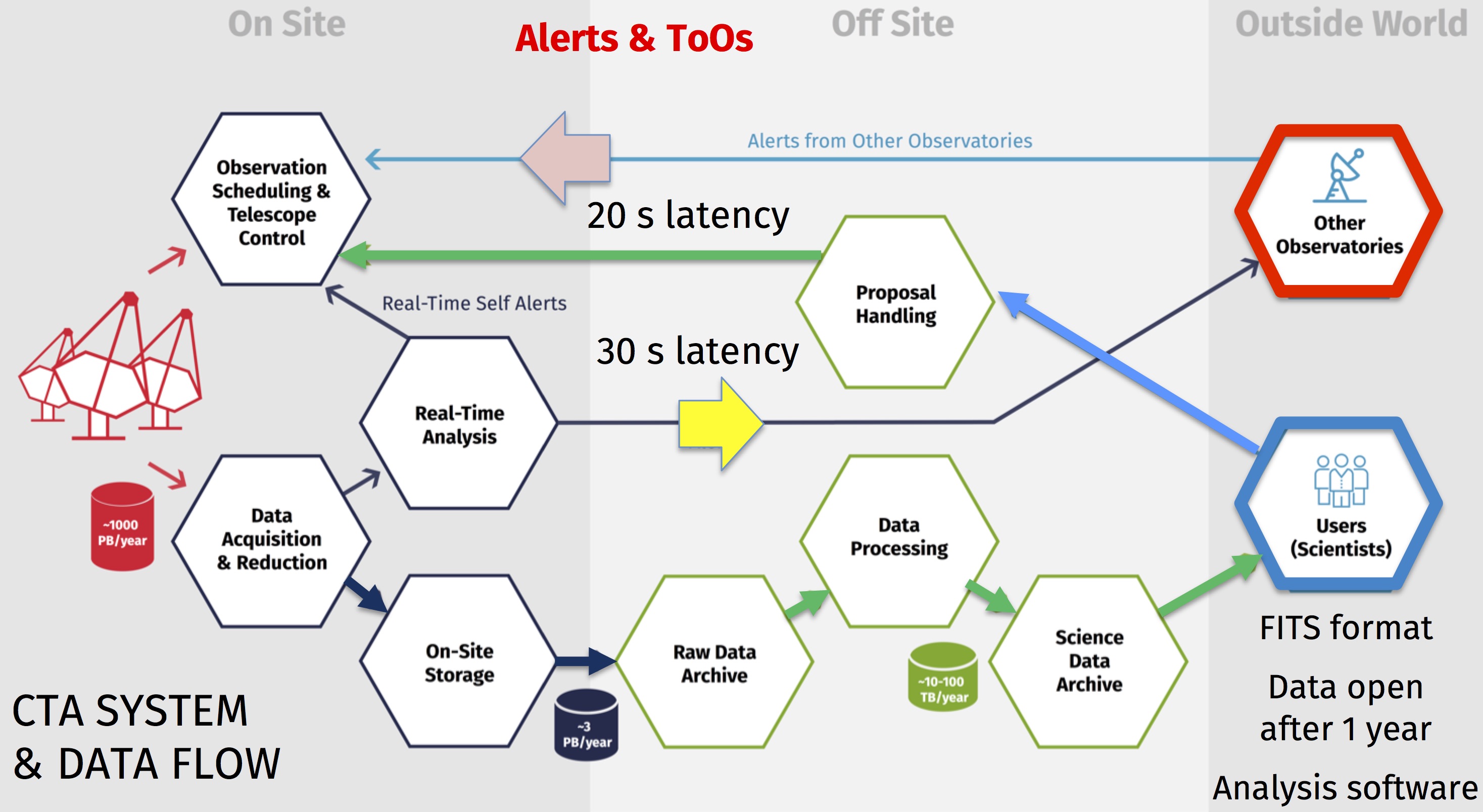}}
\caption[]{
CTA data flow.
Data are generated by the cameras at a volume of $\sim 1000$~PB/year, are reduced by
the ``Data Acquisition \& Reduction'' system, and stored in the ``On-Site Storage'' with a volume of
$\sim 3$~PB/year. At the same time, a ``Real-Time Analysis'' system analyses the data stream, issuing
alerts for flaring or bursting TeV sources to ``Other Observatories'' with a latency of 30 s.
The alerts may also change the scheduling of CTA observations via the ``Observation Scheduling \&
Telescope Control'' system.
This system also digests alerts from ``Other Observatories'' with a latency of 20 s.
Data are transferred off site and stored in a ``Raw Data Archive'', from where they are read
and processed, resulting in a volume of $\sim 10-100$~TB/year.
These processed data, that will be useful for science analysis, will be stored in the ``Science Data Archive''
that will be the access point to the CTA ``Users''.
``Users'' may also send-in observing proposals that will be handled by a dedicated ``Proposal Handling''
system.}
\label{fig:dataflow}
\end{figure}

An important aspect of CTA is its capability to react to alerts.
The TeV sky can be extremely variable, and some CTA science relies on the rapid reaction of the
observatory to triggers of flaring sources or bursts.
The CTA Observatory will be capable of both receiving alerts from external observatories and issuing
alerts during observations.
The latter will be achieved by a dedicated Real-Time Analysis system.\cite{bulgarelli2015}
Figure \ref{fig:dataflow} illustrates the CTA data flow and shows the different trigger paths, indicating
the required latencies for external triggering of observations, and the Real-Time Analysis system
sending out triggers to other observatories.

A novelty in the domain of TeV astronomy is the fact that CTA will be a proposal-driven open
observatory.
This means that upon regular Announcements of Opportunity (AO), the CTA Observatory will solicit
scientific proposals for observations to be carried out with the CTA arrays.
Data will be proprietary to the Principal Investigator (PI) of a successful observing proposal for a
period of time (typically one year after the delivery of the data to the PI), and after that period
the data will become publicly available in a CTA science data archive.
Science data management will be performed by a dedicated centre, the Science Data
Management Centre (SDMC), situated on the DESY campus in Zeuthen, Germany.
The management of the entire observatory, including maintenance and upgrades of the telescopes
and cameras, will be done by the Headquarters (HQ), situated in Bologna, Italy.

All together, HQ, SDMC and the two array sites form CTAO, the CTA Observatory, a legal
entity that is responsible for the construction and operations of CTA.
Currently, CTAO is implemented as a gGmbH under German law located in Heidelberg, Germany,
where the former Project Office (PO) of the CTA construction project was located.
After the selection of Bologna as the host site of the CTA Headquarters in 2016, the PO moved to
Bologna.
Currently, an ERIC, a European Research Infrastructure Consortium, is under preparation to serve
as the final legal entity of CTAO which is able to engage the construction of the arrays.
Hence, the creation of the ERIC by the European Commission is an important milestone, and will mark
the official start of CTA construction.
It is currently expected that this milestone will be reached in 2021.

Besides CTAO there is CTAC, the CTA Consortium of scientists and engineers who devised the CTA
concept more than a decade ago and have been the driving force behind its design.
Today, the CTA Consortium includes 1500 members from more than 200 institutes in 31 countries.
CTAC has developed and detailed CTA's key science goals \cite{ctascience} and will be responsible
for the scientific analysis and publication of scientific results of the Key Science Projects (KSPs).
The KSPs are large programmes that ensure that some of the key science issues for CTA are addressed
in a coherent fashion, with a well-defined strategy.
According to current plans, KSPs cover approximately $40\%$ of CTA's observing time.
Similar to all other data acquired by CTA, the KSP data will become public after a proprietary period.

Scientists and engineers of CTAC are also currently involved in developing hardware and software
prototypes for CTA.
All of the prototyping activities are performed in the context of CTAC, while prototypes are generally
developed by sub-consortia that have their internal organisation, structures and funding sources.
Many of the prototyping activities are nearing completion.
Consequently, CTAC is currently evolving towards a structure that is able to support CTAO with its
knowledge and expertise in the construction and operations of CTA.
In this context, a new Memorandum of Understanding has been signed between CTAC member
institutes, and new CTAC membership rules have been established that are based on commitments of
individuals to CTAC duties.

\section{Realising CTA}

\begin{figure}
\centerline{\includegraphics[width=\linewidth]{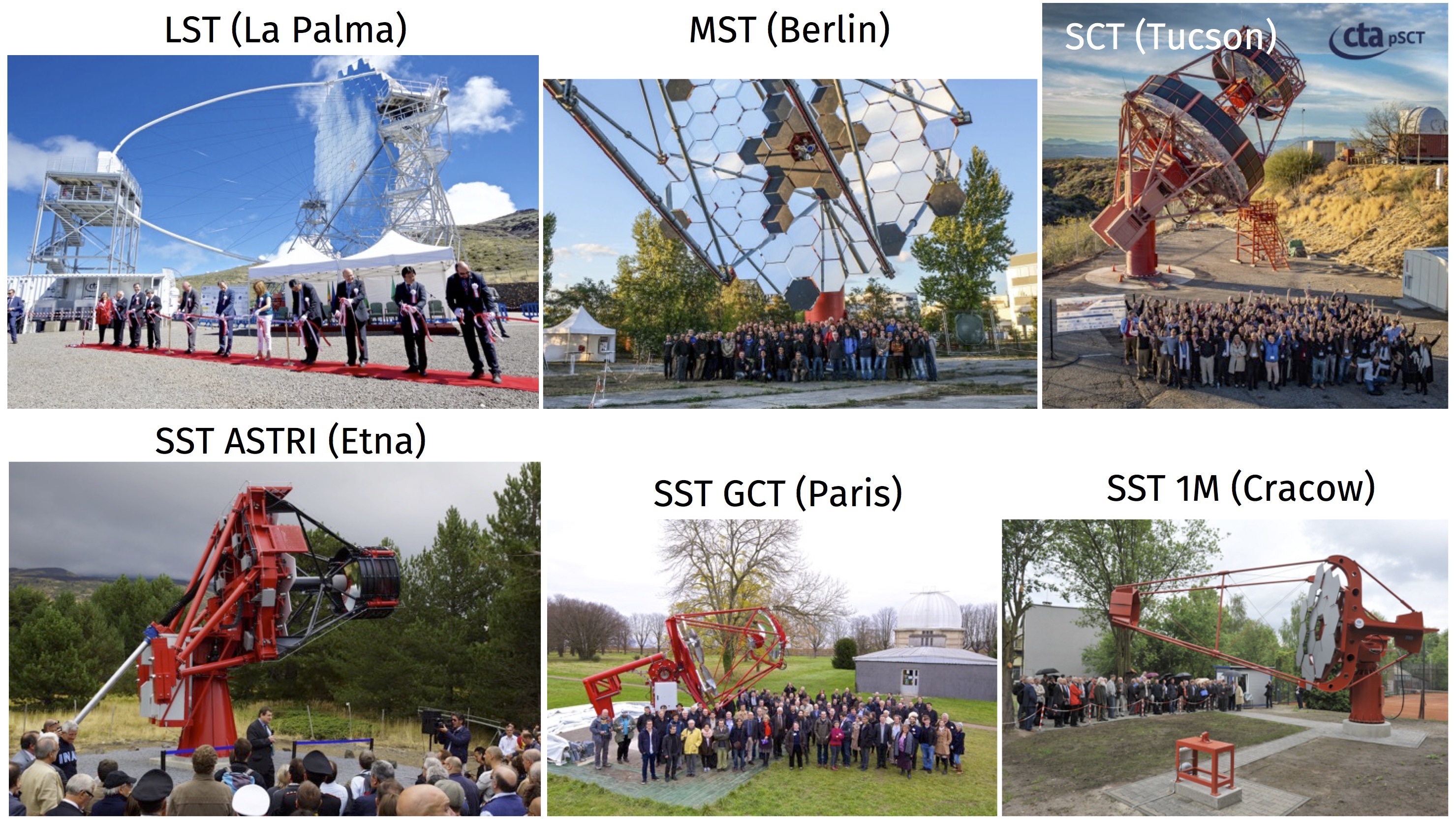}}
\caption[]{
Inaugurations of CTA prototype telescopes.
From top-left to bottom-right:
LST prototype at the Observatorio del Roque de los Muchachos, La Palma, Spain (also the site of MAGIC);
MST prototype in Berlin Adlershof, Germany;
SCT prototype at the Fred Lawrence Whipple Observatory, Tucson, USA (also the site of VERITAS);
SST ASTRI prototype at the Serra La Nave Observatory on mount Etna, Italy;
SST GCT prototype in Meudon, France;
SST 1M prototype in Cracow, Poland.}
\label{fig:prototypes}
\end{figure}

CTA will be composed of more than 100 telescopes, hence careful prototyping of all array
elements is required to assure the seamless and cost-efficient production, installation, operations
and maintenance of all hardware and software.
Integration of more than 100 telescopes into a single observatory has never been done before,
and consequently presents a major challenge to the CTA project.

Over the past decade, CTAC members have developed in total six telescope and seven
camera prototypes (see Fig.~\ref{fig:prototypes}).
All prototypes have undergone extensive testing and verifications to assure that the instruments
are working as expected and that the CTA requirements are met.
While three of the telescope prototypes are located close to the labs who developed them
(MST, SST GCT and SST 1M), the three other prototypes are installed at a site that allows for
astronomical observations (LST, SCT and SST ASTRI).
The LST and the SST ASTRI prototypes have in fact already detected with the Crab Nebula their
first TeV gamma-ray source.\cite{lombardi2020}
So to some extent, CTA had already its first light!

Nevertheless, there is still some way to go before CTA will become reality.
Firstly, CTA only requires 3 telescope types while 6 telescope prototypes exist, hence some choices
need to be made.
A first choice that was made was the selection of a Schwarzschild-Couder configuration for the SST
design based on the ASTRI telescope and CHEC camera prototypes, taking into account the experience
gained from all SST designs.
Another choice is to be made between the MST and SCT telescope designs, where SCT stands
for ``Schwarzschild-Couder Telescope'', which is an alternative implementation of a Mid-Sized Telescope.
Two camera prototypes exist for the MST, the FlashCam and the NectarCAM, and if the
MST is selected, a plausible option would be to use both cameras, one for the northern site and the
other for the southern site.

For the LST there exists no other design, and the current prototype is actually already installed on
the location of LST-1 on the northern site of CTA.
Due to the size and the related investment cost of building a LST, it was decided to build the LST
prototype on its final location, and to aim for an integration of the prototype into the final CTA
array.
So while formally the LST on La Palma is still a prototype, once fully commissioned and accepted by
CTAO, it will become CTA's LST-1.
The LST team is well on the way to achieve this goal, and the Critical Design Review that was 
organised in late 2019 has confirmed that there are no show stoppers on the path.

\begin{figure}
\centerline{\includegraphics[width=\linewidth]{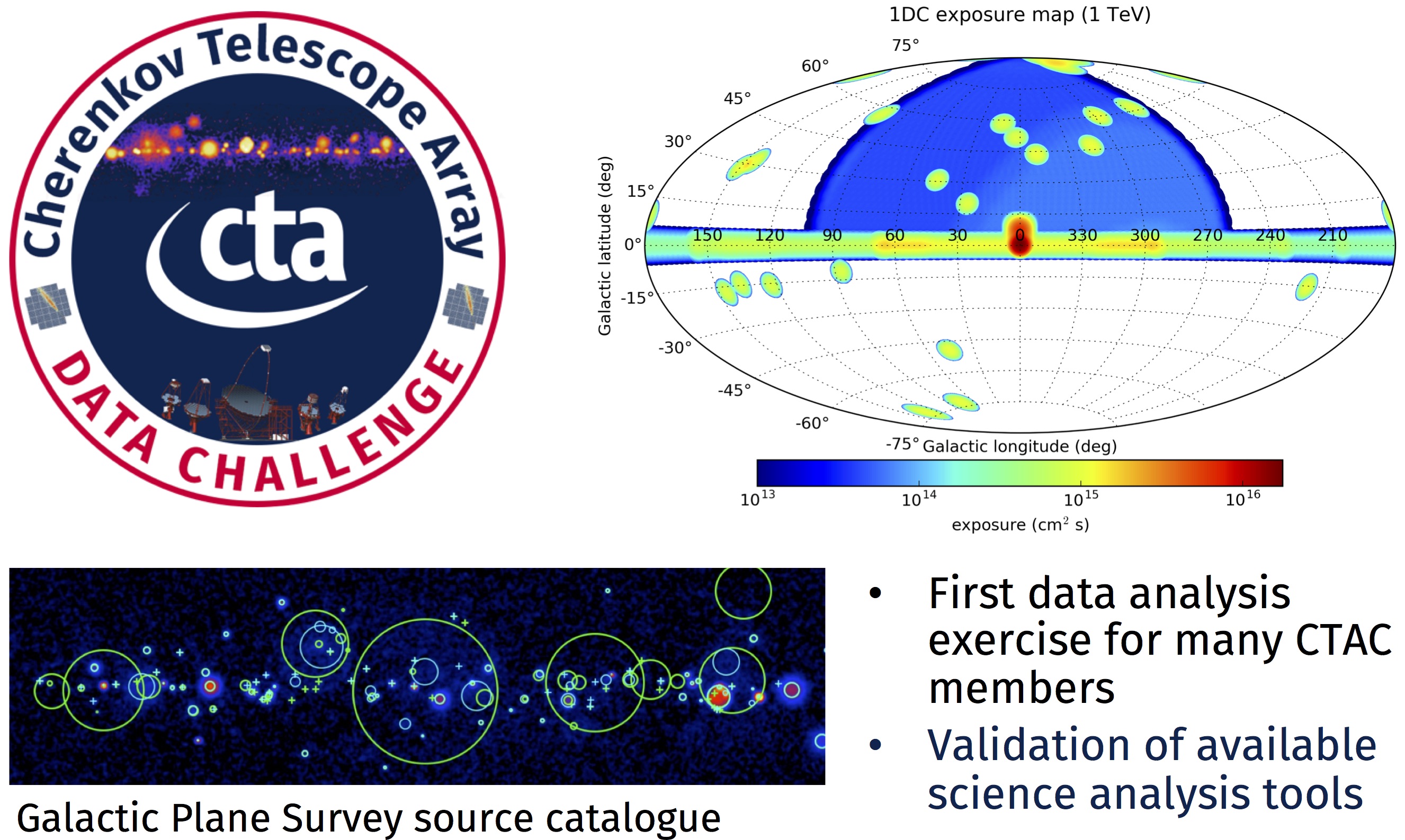}}
\caption[]{
First Data Challenge.
The top-right panel illustrates the exposure of the simulated data, comprising
a survey of the Galactic plane (1635 hours),
a survey of the Galactic centre (836 hours),
an extragalactic survey (530 hours), and
an AGN monitoring program (960 hours).
The bottom-left panel illustrates the results of a catalogue pipeline applied to the
data of the Galactic plane survey.\cite{cardenzana2018}
}
\label{fig:1dc}
\end{figure}

But CTA is not only hardware.
Software is equally important for an IACT, and needed to convert the images of Cherenkov
light registered by the cameras into gamma-ray event candidates with estimated photon arrival
directions and energies (see Fig.~\ref{fig:dataflow}).
CTAC members are currently actively prototyping analysis chains for the processing of CTA
data,\cite{kosack2019,pintore2020}
and some of these chains were used to achieve the first light detections of the Crab nebula
by the LST and ASTRI SST teams.

Another challenge is related to the fact that CTA will be an open observatory, hence non-expert
users need to be able to analyse the data products.
CTA will distribute to each successful PI the data and corresponding instrument response functions
in FITS format, along with science analysis software that allows the extraction of images, spectra and light
curves from the data.\cite{knoedlseder2015}
Also here CTAC members are actively prototyping software,\cite{knoedlseder2016,deil2017}
and CTAC organised in 2017 an extensive Data Challenge to test this software on several
years of simulated KSP data.
Figure \ref{fig:1dc} illustrates the simulated data products and the results of a catalogue pipeline
applied to the simulated data of the Galactic plane survey.\cite{cardenzana2018}
For many CTAC members this was the first time that they were in contact with prototype CTA
data, and the exercise demonstrated that the science analysis software prototypes are working
seamlessly, allowing non-expert users to extract results from the data.
More recently, the software was also exercised on real data obtained with the H.E.S.S. telescopes,
confirming their readiness for CTA science analysis.\cite{knoedlseder2019,mohrmann2019,nigro2019}

\section{Conclusion}

I have not yet touched in this contribution on the science potential of CTA.
A detailed description of the CTA science case and specifically the Key Science Projects that are
proposed by CTAC can be found in the ``Science with CTA'' book.\cite{ctascience}
In summary, CTA will contribute to understanding cosmic particle acceleration, unveiling where and how
particles are accelerated, how the accelerated particles propagate, and how they impact their
environment.
CTA will probe extreme environments, such as those found close to neutron stars and black holes
where magnetic and gravitational energy is transferred to relativistic particles.
CTA will study the physics of relativistic jets, winds and explosions.
CTA will use distant gamma-ray sources as beacons to study the intergalactic medium along
the line of sight, probing the cosmic background light density as function of redshift and the strength
of the intergalactic magnetic field.
CTA will also explore the frontiers of physics, searching for annihilation signatures of elusive dark matter
particles and testing the fundamental principle of Lorentz invariance with unprecedented precision.

Much of the science that CTA will do requires input from a wider community, since results need to be 
considered in a multi-wavelength and multi-messenger context, and obviously need to be confronted to models
and theories.
Having in this case an open and proposal driven observatory is the most logical and efficient thing
to do.
Around $50\%$ of CTA's observing time will be available to the wider community, hence the success
of CTA will strongly depend on the engagement of the community in CTA science.
All CTA data will become available in a public archive after a proprietary period of typically one year, and
mining this archive will be a complementary and important way of doing science with CTA.

The deployment of CTA is now quickly approaching, and first science data will become available
soon.
Hence now it is time to engage with CTA, to familiarise with the data and the analysis tools, and to
prepare for sending in excellent observation proposals that will contribute to the exploration of the
high-energy Universe beyond its established horizons.

\section*{Acknowledgments}

We gratefully acknowledge financial support from the agencies and organizations listed here:
\url{http://www.cta-observatory.org/consortium_acknowledgments}.
This paper has gone through internal review by the CTA Consortium.


\end{document}